\documentstyle[12pt]{article}
\input{amssym.def}
\input{amssym.tex}

\def\thebibliography#1{\list
 {$^{\arabic{enumi}}$}{\settowidth\labelwidth{[#1]}\leftmargin\labelwidth
 \advance\leftmargin\labelsep
 \usecounter{enumi}}
 \def\newblock{\hskip .11em plus .33em minus .07em}
 \sloppy\clubpenalty4000\widowpenalty4000
 \sfcode`\.=1000\relax}

\begin{document}

\makeatletter
\renewcommand{\@cite}[2]{{\footnotesize $^{#1\if@tempswa , #2\fi}$}}
\def\ncite#1{{\def\@cite##1##2{##1}\cite{#1}}}
\makeatother

\baselineskip=.7cm

{\large \noindent {\bf DISTORTION OF A PHASE SPACE UNDER

\vspace{3mm}\noindent THE DARBOUX TRANSFORMATION}\footnote{
To appear in J. Math. Phys. 1998, {\bf 39}, No 1 or No 2.}}
\vspace{5mm}
\begin{quotation}
\noindent Boris F. Samsonov\footnote{Electronic mail address: samsonov@phys.tsu.tomsk.su}

\noindent {\it Tomsk State University, 36 Lenin Avenue, 634050 Tomsk, Russia,
}

\vspace{5mm}
\noindent The Darboux transformation operator technique is applied to construct
exactly solvable anharmonic singular oscillator potentials and to study
their coherent states. Classical system corresponding to a transformed quantum system is constructed with the help of the coherent
states technique. It is shown that at classical level the Darboux
transformation may be treated as a transformation of K\"ahler potential
which leads to a distortion of the initial phase space.

\noindent PACS: 03.20.+i; 03.65.-w; 03.65.Fd; 03.65.Ge
\end{quotation}

\hspace{10mm}

\begin{flushleft}
{\bf I. INTRODUCTION}
\end{flushleft}
\hspace{5mm}

In quantum mechanics there exists a beautiful recipe to construct a family
of
exactly solvable Hamiltonians starting from an initial Hamiltonian $h_0$
for which
the solutions of the Schr\"odinger equation $h_0\psi _E=E\psi _E$ are known.
Since this method is very simple in applications and possesses an
intrinsic charm
it attracts a thorough attention of several generations of researches.
A central point of this theory is a notion of a transformation
operator
introduced and investigated by Delsart and Lions.\cite{Delsart}
According to their approach, every operator $L$ intertwining the initial
Hamiltonian $h_0$ and transformed Hamiltonian $h_1$,
\begin{equation}\label{itert}
Lh_0=h_1L,
\end{equation}
is a
{\it transformation operator} (originally called a transmutation operator\cite{Delsart}). It is obvious that $L$
transforms solutions $\psi _E$ ($\psi _E\notin {\rm ker}L$)
of the initial Schr\"odinger equation onto solutions $\varphi _E=L\psi _E$ of a
{\it transformed Schr\"odinger equation} $h_1\varphi _E=E\varphi _E$
corresponding to the same eigenvalue $E$. In the case of  a first order differential operator $L$
and a one dimensional Hamiltonian $h_0$, the
relation (\ref{itert}) leads to\cite{BSDar} the transformation proposed by Moutard\cite{Mout} and studied by Darboux \cite{Darb} (see also Ref.\ncite{Ince}). If $L$ is a differential operator of order $N$ it is
called\cite{BSDar} {\it $N$-order Darboux transformation operator}. In this paper we will restrict ourselves by first order transformations.

The Darboux transformations operator technique is a very useful tool
in obtaining and investigating new exactly solvable quantum problems.
It enters many fields of physics and mathematics.
In soliton theory the Darboux transformation permits one to construct and
investigate properties of multisoliton solutions to nonlinear
evolution equations (see Ref.\ncite{Matv} and references therein).
In quantum mechanics this technique takes the form of a factorization
method. Using this method Schr\"odinger\cite{Schr} was able to solve the hydrogen
atom problem in a pure algebraic manner.
Subsequently, Infeld and Hall\cite{InfHall}
generalized the Schr\"odinger method and obtained a wide
class of solvable potentials.
In the inverse quantum scattering problem\cite{FadZS} this transformation is used to
construct potentials with given properties and to obtain appropriate solutions to the Schr\"odinger equation.
New developments in this field have recently been presented in
Ref.\ncite{SuzkoSam}.
 In the context of supersymmetric quantum mechanics (for reviews see Ref.\ncite{SUSY}) the
Hamiltonian $h_0$ and its superpartner $h_1$ form a matrix-valued
Hamiltonian (so called super Hamiltonian) ${\cal H}$. Darboux transformation
operator $L$ and its Hermitian conjugate $L^+$ participate in construction
of supercharge operators which, together with ${\cal H}$, form a superalgebra. Initially the supersymmetric quantum mechanics served as a toy model to illustrate the problem of supersymmetry breaking in quantum field theories. Recently,  new features in s
upersymmetry breakdown in quantum mechanics caused by higher order Darboux transformations have been discovered.\cite{SamMPLA1} Besides quantum field theories the same ideas find applications in many others fields of theoretical physics such as nuclear, a
tomic, solid state and statistical physics.\cite{SUSY}

\makeatletter
\renewcommand{\@oddhead}
{{\footnotesize
\hfill
Boris F. Samsonov: Distortion of a phase space \hfill }}
\makeatother

In spite of the long period of application of the Darboux transformation
operator method to quantum mechanical problems the question of
correspondence between the transformed quantum
system and a classical system has not been properly discussed in the literature.
In a recent paper\cite{NFilho97} a classical mechanics counterpart of the transformed system has been described. It consists in addition of a total time derivative of a purely imaginary function to the Lagrangian of a classical system. This transformation
 does not affect the equations of motion. Canonical quantization of the new classical system gives the transformed quantum system.

The concept of coherent states plays an important role in several fields of
physics and mathematics (see, for example, Refs.\ncite{Per}-\ncite{MM}). In this
paper we exploit the property of coherent states to establish the
correspondence between the quantum mechanics and the classical mechanics.
For any quantum system we can construct using
the coherent states technique\cite{Per,Klauder} a classical system possessing the following remarkable
property. The geometric quantization\cite{JRK} of the resulting
system gives holomorphic representation of the initial quantum system. By
these means we may establish a correspondence between quantum and
classical systems.
Recently it has been noted\cite{BS} that application of the Darboux transformation operator to the coherent states of the initial quantum system gives a set of states which may be treated as coherent states of the transformed system. This observation open
s the door to establishing a correspondence between transformed quantum and classical systems and obtaining a classical counterpart of the Darboux transformation.

In this paper we realize the above program for a system with the Hamiltonian
\begin{equation}
\label{e1}h_0=-d^2/dx^2+x^2/4+b/x^2,\quad b>0,\quad x\in [0,\infty ).
\end{equation}
We will not discuss the significance of the spectral problem for the
Hamiltonian (\ref{e1}) (more precisely, for its closure in the Hilbert space $%
H $ of square integrable on interval $[0,\infty )$ functions with the zero
boundary condition at the origin). We only notice the fact that this Hamiltonian
appears in different physical problems such as central potential problem,
Calogero problem\cite{Cal}, fractional statistics and anyons\cite{LM},
two-dimensional QCD\cite{MP}, and others.

We obtain exactly solvable partners for the Hamiltonian (\ref{e1}). Next
we construct the coherent states for the transformed $a$ $l\acute a$ Darboux
Hamiltonians. These states as well as the coherent states for the input system (initial coherent states) are
labeled by the points of a unite disk on the complex $z$-plane. We consider two different K\"ahler structures on the unit disk. The first one is the well known
Poincar\'e model of the Lobachevsky plane and the second is the Darboux
transformation of the first. The
expression for the transformed K\"ahler potential is obtained. We also notice that in contrast to the initial manifold
the transformed one has a varying
curvature. The integration measure realizing the identity decomposition for the
transformed coherent states is calculated. The holomorphic representation
for the discrete basis set as well as for the coherent states, lowering and
raising operators and Darboux transformation operators is constructed.
It is established that at the classical level the Darboux transformation may be considered as such a transformation of the K\"ahler potential and Hamilton function that preserves the equations of motion.

\hspace{5mm}
\begin{flushleft}
{\bf II. COHERENT STATES OF SINGULAR OSCILLATOR}
\end{flushleft}

\hspace{5mm}
In this section we briefly review the well known\cite{Per} properties of
the coherent states for the Hamiltonian (\ref{e1}) that we will need further.

The dynamical symmetry algebra for the quantum system with the Hamiltonian (\ref
{e1}) is $su(1,1)$. In coordinate
representation the generators of this algebra are expressed in terms of the harmonic oscillator annihilation $a$
and creation $a^{+}$ operators
$$
\begin{array}{c}
k_0=\frac {1}{2}h_0,\quad k_{+}=\frac {1}{2}\left[
(a^{+})^2-b/x^2\right] ,\quad k_{-}=\frac {1}{2}\left[
a^2-b/x^2\right] ,\quad  \\ a=d/dx+x/2,\quad a^{+}=-d/dx+x/2.
\end{array}
$$
and satisfy the standard commutation relations
$$\left[ k_0,k_{\pm }\right] =
\pm k_{\pm },\quad \left[ k_{-},k_{+}\right] =2k_0\ .$$
 Since we
consider an irreducible representation the corresponding $su(1,1)$
Casimir operator takes the
constant value ${\cal C}=\frac 12\left[ k_{+}k_{-}+k_{-}k_{+}\right]
-k_0^2=3/16-b/4=k(1-k)$. The value of $k=1/2+(1/4)\sqrt{1+4b}$ defines the
ground state (vacuum) energy $E_0=2k$. The vacuum state $|0\rangle $ is
defined by the equations: $k_{-}\mid 0\rangle =0$ and $k_0\mid 0\rangle =k\mid
0\rangle $. Other discrete basis eigenfunctions $\mid n\rangle $ of the
Hamiltonian $h_0$ are defined by means of the raising operator $k_{+}$
$$
k_{+}\mid n\rangle =-\sqrt{(n+1)(n+2k)}\mid n+1\rangle ,
$$
$$
\mid n\rangle =(-1)^n\sqrt{\frac{\Gamma (2k)}{n!\Gamma (2k+n)}}\left(
k_{+}\right) ^n\mid 0\rangle ,\quad k_0\mid n\rangle =(k+n)\mid n\rangle
$$
and in the coordinate representation they have the form
\begin{equation}
\label{e2}\psi _n(x)=\langle x\mid n\rangle =\left[ n!2^{1-2k}\Gamma
^{-1}(n+2k)\right] ^{1/2}x^{2k-1/2}\exp (-x^2/4)L_n^{2k-1}(x^2/2),
\end{equation}
where $L_n^\alpha (z)$ are the Laguerre polynomials.

The coherent states $\mid z\rangle $ can be obtained by applying of the unitary
displacement operator $D_z$ to the vacuum state
$$
\mid z\rangle =D_z\mid 0\rangle =e^{zk_{+}}\exp \left[ \ln (1-z\overline{z}%
)k_0\right] e^{-\overline{z}k_{-}}\mid 0\rangle =N_{0z}\sum_{n=0}^\infty
a_nz^n\mid n\rangle ,
$$
$$
N_{0z}=(1-z\overline{z})^k,\quad a_n=(-1)^n\sqrt{\frac{\Gamma (2k+n)}{%
n!\Gamma (2k)}},\quad \left| z\right| <1.
$$
In the coordinate representation they look like
\begin{equation}
\label{e3}
\begin{array}{c}
\psi _z(x)=2^{1/2-k}\Gamma ^{-1/2}(2k)(1-z)^{-2k}(1-z
\overline{z})^kx^{2k-1/2}\times \\ \exp \left[
-(1/4)(1-z)^{-1}(1+z)x^2\right] ,\quad \left| z\right| <1.
\end{array}
\end{equation}

The linear manifold spanned by the vectors $\{\mid z\rangle \}$ forms an everywhere dense set in the
Hilbert space $H$ with the following identity
decomposition:
\begin{equation}
\label{e4}\int _{\left| z\right| <1}\mid z\rangle \langle z\mid d\mu
(z)=1, \quad d\mu (z)=\frac{2k-1}\pi (1-z\overline{z})^{-2}dzd\overline{z}.
\end{equation}
The system $\{\mid z\rangle \}$ is overcomlete.
The completeness of different subsystems of this system was studied in
detail in Ref.\ncite{per}.

The Fourier coefficients $c_n$ of a vector $\mid \psi \rangle \in H$ with
respect to the basis $\{\mid n\rangle \}$ define a holomorphic
representation $\psi (z)$ of the vector $\mid \psi \rangle $ in the space of functions which are holomorphic in the unit
disk
$$
\langle \overline{z}\mid \psi \rangle =N_{0z}\psi \left( z\right) ,\quad
\psi \left( z\right) =\sum\limits_{n=0}^\infty a_nc_nz^n.
$$
The functions $\psi _n(z)=a_nz^n$ realize the holomorphic representation of
the basis vectors $\mid n\rangle $. For the holomorphic representation of the
coherent state vectors we have
$$
\zeta ^{\left( 0\right) }\left( z\right) =N_{0z}^{-1}\langle \overline{z}%
\mid \zeta \rangle =\left( 1-\zeta \overline{\zeta }\right) ^k\left( 1-\zeta
z\right) ^{-2k}.
$$

Let ${\cal L}$ be the lineal of entire	analytic functions
$\psi(z)$ defined in the unit disk such that
$$\int _{|z|<1}\mid \psi (z)\mid ^2(1-|z|^2)^{2k}d\mu (z)<\infty .$$
If now we define an inner product in the lineal ${\cal L}$
\begin{equation}
\label{sp}\langle \psi _1\left( z\right) \mid \psi _2\left( z\right) \rangle
\equiv \int _{|z|<1}e^{-f^{(0)}}\overline{\psi }_1\left( z\right) \psi _2\left(
z\right) d\mu \left( z\right) =\langle \psi _1\mid \psi _2\rangle ,
\end{equation}
where $\langle \psi _1\mid \psi _2\rangle $ is the inner product in the
Hilbert space $H$ then the completion of ${\cal L}$ with respect to this
inner product gives the Hilbert space ${\cal H}$. The function $f^{\left(
0\right) }=\ln \left| \langle 0\mid z\rangle \right| ^{-2}=\ln (1-z\overline{%
z})^{-2k}$ is the K\"ahler potential in the unit disk. The corresponding metric
has the form
$$
ds^2=gdzd\overline{z},\quad g=\frac{\partial ^2f^{\left( 0\right) }}{%
\partial z\partial \overline{z}}=\frac{2k}{\left( 1-z\overline{z}\right) ^2}%
.
$$

As usually the imaginary part of this metric defines a symplectic
2-form $\omega =-igdz\wedge d\overline{z}$ and consequently a Poisson
bracket $\{H_1,H_2\}$ of the functions $H_1$ and $H_2$ defined on the unit disk. The
unit disk represents now a phase space of a constant Gauss curvature ${\cal K}%
^{(0)}=-\frac 2k$. Making use of the covariant Berezin symbols\cite{Ber} we
establish a one-to-one correspondence between the classical observables and the
holomorphic representation of the quantum mechanical operators. For example, the
following classical observables:
$$
K_0=\langle \overline{z}\left| k_0\right| \overline{z}\rangle =k\frac{1+z%
\overline{z}}{1-z\overline{z}},
$$
$$
K_{-}=\langle \overline{z}\left| k_{-}\right| \overline{z}\rangle =\frac{2k%
\overline{z}}{1-z\overline{z}},\quad K_{+}=\langle \overline{z}\left|
k_{+}\right| \overline{z}\rangle =\frac{2kz}{1-z\overline{z}}
$$
correspond to the generators of $su(1,1)$ algebra of quantum operators. They
form the basis of the algebra isomorphic to $su(1,1)$ with respect to the introduced Poisson
bracket.

If we choose $K_0$ as a classical Hamilton function then the evolution of
the system is described by the Hamilton equations
$$
\dot z=\left\{ z,K_0\right\} =-iz,\quad \dot{\overline{z}}=\left\{ \overline{%
z},K_0 \right\} =i\overline{z},
$$
where the dot stands for the time derivative.

Geometric quantization\cite{JRK} of the obtained classical system%
\footnote{
In our case this is a quantization on the Lobachevsky plane\cite{Per} which really is a Berezin\cite{BerQuant} quantization on a K\"ahler manifold} leads to the
holomorphic representation of the quantum system described above. The value $h=\left(
2k\right) ^{-1}$ plays the role of the Plank constant.\cite{Per}

To conclude this section we present the explicit expressions for the Bergman kernel
which takes off the integration in the ${\cal L}$-space
$$
\delta ^{\left( 0\right) }\left( z,\overline{z}^{\prime }\right)
=\sum_{n=0}^\infty \psi _n\left( z\right) \overline{\psi }_n\left( z^{\prime
}\right) =\left( 1-z\overline{z}^{\prime }\right) ^{-2k},
$$
and for the $su(1,1)$ generators
$$
k_0\left( z\right) =z\frac d{dz}+k,\quad k_{+}\left( z\right) =z^2\frac
d{dz}+2kz,\quad k_{-}\left( z\right) =\frac d{dz}.
$$
in holomorphic representation.
\vspace{5mm}

\begin{flushleft}
{\bf III. DARBOUX TRANSFORMATION OF COHERENT STATES}
\end{flushleft}

\vspace{5mm}
In a simplest case\cite{Ince,BSDar} the Darboux transformation operator is the first
order differential operator of the form
\begin{equation}
\label{L} {L}=-L_0(x)+d/{dx}=-u^{\prime }(x)/u(x)+d/{dx}
\end{equation}
where the prime denotes the derivative with respect to $x$. When acting on the solutions $\psi _n(x)$ of the initial Schr\"odinger
equation
\begin{equation}
\label{h0}h_0\psi _n(x)=E_n\psi _n(x),\quad E_n=2n+2k
\end{equation}
it transforms them into the solutions of another Schr\"odinger equation
$$
\varphi _n(x)=N_n{L}\psi _n(x),\quad h_1\varphi _n(x)=E_n\varphi
_n(x)
$$
with the same eigenvalues $E_n$. The factor $N_n$ is introduced to ensure that the states $\varphi _n$ are normalized to unity. New exactly solvable Hamiltonian $%
h_1=h_0+A(x)$ is defined by the potential difference $A=A(x)=-2(\ln
u)^{\prime \prime }$. The function $u=u(x)$ called {\it transformation function}%
\cite{BSDar}
is a solution to the initial Schr\"odinger equation
$$
h_0u(x)=\alpha u(x)
$$
with $\alpha \leqslant E_0$.

For the sake of simplicity we will restrict ourselves to the case
when $\alpha <E_0$, $u(x)\ne 0$ $\forall x\in
(0,\infty )$ and $1/u(x)$ is not a square integrable function in the interval
$[0,\infty )$. In this  case $u\notin H$ and the set $\{\mid \varphi _n\rangle \}$
constitutes a complete basis in the Hilbert space $H$ provided the initial
system $\{\mid \psi _n\rangle \}$ is complete.\cite{BSDar} In terms of the supersymmetric quantum mechanics this case corresponds to a broken supersymmetry.

Let us now choose the following solution of the initial equation (\ref{h0}):
$$
u=u_p(x)=x^{2k-1/2}L_p^{2k-1}\left( y\right) \exp \left( x^2/4\right) ,\quad
y=-x^2/2 ,
$$
$$
\alpha =-2(k+p),\quad p=0,1,2,\ldots
$$
as the transformation function. It is worth stressing that
$L_p^{2k-1}(y)\ne 0$ $\forall x\ne 0$ when
$y=-x^2/2$$(<0)$ and, hence, the function $u_p(x)$ is nodeless for $x>0$.
The function $L_0(x)=L_{0p}(x)$ has in this case
the form
$$
L_{0p}(x)=\frac{1-4k}{2x}-\frac x2-\frac{xL_{p-1}^{2k}\left( y\right) }{%
L_p^{2k-1}\left( y\right) }
$$
and for the potential difference we obtain
$$
A=A_p\left( x\right) =-1+\frac{4k-1}{x^2}-
$$
$$
2\frac{x^2L_{p-2}^{2k+1}\left( y\right) +L_{p-1}^{2k}\left( y\right) }{%
L_p^{2k-1}\left( y\right) }+2x^2\left[ \frac{L_{p-1}^{2k}\left( y\right) }{%
L_p^{2k-1}\left( y\right) }\right] ^2, \quad y=-x^2/2.
$$
It is worth noticing that because of the condition $L_p^{2k-1}(y)\ne 0$ for $x\ne 0$ the
potential difference $A_p(x)$ has a single pole at $x=0$. The asymptotic
 behavior of this function near the point $x=0$ is as follows: $A_p(x) \rightarrow
-1-p/k+(4k-1)/x^2$. In the limit as $x\rightarrow \infty $ this function tends to $-1$.
For $b\ne 0$ the behavior of the transformed potentials
$V_p(x)=V_{0}(x)+A_p(x)$ is nearly the same as
the behavior of the initial potential $V_{0}(x)=x^2/4+b/x^2$.
There is only a small difference
in the region of the minimum. When $b=0$ the initial potential is regular at $x=0$ whereas
all $V_p(x)$ are singular: $V_p(x)\rightarrow -1-4p/3+2/x^2$. As an illustration, we
give the explicit expression for two first potential differences
$$A_0(x)=-1+\frac{4k-1}{x^2},\quad
A_1(x)=A_0(x)+\frac 4{4k+x^2}-\frac{32k}{\left( 4k+x^2\right) ^2}\ . $$
Thus, the transformed Hamiltonians $h_1=h_1^{(p)}=-d^2/dx^2+V_p(x)$
are bounded from below and are essentially self adjoint operators defined
on the dense set in the Hilbert space $H$.
All these Hamiltonians have only discrete spectrum.
The position of the energy levels is the same as before transformation:
$E_n=2n+2k$, $n=0,1,2,\ldots $.

The operator ${L}^{+}=-L_0(x)-d/dx$ realizes the transformation in the
inverse direction
\begin{equation}
\label{psin}|\psi _n\rangle =N_n {L}^{+}|\varphi _n\rangle ,
\end{equation}
and together with ${L}$ participates in the following factorization:
\begin{equation}
\label{fac}{L}^{+}{L}=h_0-\alpha ,\quad {L}{L}^{+}=h_1-\alpha \ .
\end{equation}
We note that in our case operators ${L}$ and ${L}^{+}$ are
well-defined $\forall \psi \in H$ and are conjugated to each other with respect to
the inner product in the space $H$. This observation together with the factorization
properties (\ref{fac}) is very useful in calculating integrals involving
transformed functions. In particular, for the normalization constant $N_n$ we readily
find $N_n^{-2}=\langle \varphi _n|h_1-\alpha |\varphi _n\rangle =E_n-\alpha
=2p+4k+2n.$

Now, by following the method of Ref.\ncite{BS} we find the transformed coherent states
$$
\varphi _z(x)=N_{1z} {L}\psi _z(x)=N\sum\limits_{n=0}^\infty
b_nz^n\varphi _n,
$$
$$
N=N_{0z}N_{1z}/N_0,\quad b_n=a_nN_0/N_n
$$
From the factorization properties (\ref{fac})
one readily derives the value of the normalization constant
$$
N_{1z}^{-2}=\frac{4k+2p-2pz\overline{z}}{1-z\overline{z}}.
$$

Using the identity operator
\begin{equation}
\label{I}\sum_{n=0}^\infty \left| \varphi _n\rangle \langle \varphi
_n\right| =1
\end{equation}
one can calculate the integration measure $\nu (z)$ which realizes the following identity
decomposition:
\begin{equation}
\label{Inu}\int _{\left| z\right| <1}\left| \varphi _z\rangle \langle
\varphi _z\right| d\nu \left( z\right) =1.
\end{equation}
For this purpose we pass from the complex variables $z$ and $\overline{z}$
to the polar ones $z=\sqrt{x}\exp {i\phi }$ and look for the measure in the form $d\nu =\frac 12h\left( x\right) dxd\phi $. The integration over the
angle variable $\phi $ in (\ref{Inu}) is trivial. Then making use of the
formula (\ref{I}), we obtain the equation for the function $h(x)$
$$
\pi \int _0^1dxh(x)\frac{x^n\left( 1-x\right) ^{2k+1}}{2k+p-px}=\frac{%
n!\Gamma \left( 2k\right) }{\Gamma \left( n+2k\right) \left( n+2k+p\right) }%
.
$$
So, we arrive to a moments problem well known in mathematics.\cite{moment} In our case the solution to this problem can readily be
obtained with the help of the integral representation of the beta-function
$$
B\left( a,b\right) =\frac{\Gamma \left( a\right) \Gamma \left( b\right) }{%
\Gamma \left( a+b\right) }=\int _0^1x^{a-1}\left( 1-x\right) ^{b-1}dx
$$
and the following identity:
$$
\frac{\Gamma \left( n+1\right) \Gamma \left( 2k\right) }{\left(
n+2k+p\right) \Gamma \left( n+2k\right) }=\sum_{j=0}^pC_p^j\frac{2k-1}{%
2k+p-j-1}B\left( n+j-1,2k+p-j\right)
$$
where $C_p^j$ are the binomial coefficients. It is not difficult to
prove this identity by induction. The final expression for the function $h(x)$ is
$$
h\left( x\right) =\frac{2k-1}\pi \left( 2k+p-px\right) \sum_{j=0}^pC_p^j%
\frac{x^j\left( 1-x\right) ^{p-j-2}}{2k+p-j-1}.
$$
Once we know the measure $\nu (z)$ we can construct the holomorphic
representation $\varphi (z)$ in the unit disk of a vector
$|\varphi \rangle \in H$ which is
defined by its Fourier coefficients $c_n$ with respect to the base $%
\{|\varphi _n\rangle \}$
$$
\langle \varphi _{\overline{z}}|\varphi \rangle =N\varphi \left(
z\right) ,\quad \varphi \left( z\right) =\sum_{n=0}^\infty b_nc_nz^n.
$$
 It follows from these relations the holomorphic representation for the transformed discrete basis vectors $%
\varphi _n(z)=b_nz^n$ and for the transformed coherent states
$$
\zeta ^{\left( 1\right) }\left( z\right) =N^{-1}\langle \varphi _{\overline{z%
}}|\varphi _\zeta \rangle =\frac{2k+p+p\zeta z}{\sqrt{2k+p}}\frac{\left(
1-\zeta \overline{\zeta }\right) ^{k+1/2}}{\left( 1-\zeta z\right) ^{2k+1}}%
\left( 2k+p-p\zeta \overline{\zeta }\right) ^{-1/2}.
$$

With the help of the transformation operators $L $ and $L^+$ we construct the lowering and raising operators $p_\pm = {L}k_\pm
{L}^+$ for the transformed discrete basis functions
$$
p_+|\varphi _n\rangle = -(N_nN_{n+1})^{-1}[(n+1)(n+2k)]^{1/2}|\varphi _{n+1}
\rangle ,%
$$
$$
p_-|\varphi _n\rangle = -(N_nN_{n-1})^{-1} \left[ n(n+2k-1)\right]
^{1/2}|\varphi _{n-1} \rangle .%
$$
Together with the operator $p_0=\frac{1}{2}h_1$ they form a non-linear algebra
$$
\left[ p_0,p_{\pm }\right] =\pm p_{\pm },\quad \left[ p_{-},p_{+}\right]
=2\left( 2k(1-k)-p_0\alpha +4p_0^2\right) \left( 2p_0-\alpha \right) .%
$$
It is interesting that the operators $p_\pm $ are of the fourth degree in derivative
in the coordinate representation while in the holomorphic one they are of the second
degree
$$
p_0\left( z\right) =z\frac d{dz}+k,\quad p_{-}\left( z\right) =2z\frac{d^2}{%
dz^2}+2\left( 2k+p\right) \frac d{dz},
$$
$$
p_{+}\left( z\right) =2z^3\frac{d^2}{dz^2}+2z\left( 2k+p+2\right) \left(
z\frac d{dz}+2k\right) .%
$$

Since the operators $ {L}$ and ${L}^{+}$ realize the
correspondence between the basis sets $\{|\psi _n\rangle \}$ and $\{|\varphi
_n\rangle \}$ we can find their holomorphic representation
$$
{L}(z)=\sqrt{\frac 2{2k+p}}\left( z\frac d{dz}+2k+p\right) ,\quad
{L}^{+}(z)=\sqrt{2(2k+p)}.
$$

By assuming that the inner product of two vectors $|\varphi _1\rangle $ and
$|\varphi _2\rangle $
is independent on the representation used and taking into account the identity
decomposition (\ref{Inu}) we can equip the lineal ${\cal L}$ with a new inner product
\begin{equation}
\label{nsc}\langle \varphi _1|\varphi _2\rangle =\langle \varphi _1\left(
z\right) |\varphi _2\left( z\right) \rangle =\int _{\left| z\right| <1}%
\overline{\varphi }_1\left( z\right) \varphi _2\left( z\right) e^{-f^{\left(
1\right) }}d\nu \left( z\right) .
\end{equation}
It is naturally to declare the function
\begin{equation}
\label{fz}f^{(1)}=f^{(1)}(z,\overline{z})=\ln \left| \langle \varphi
_0|\varphi _z\rangle \right| ^{-2}=f^{(0)}+\ln \frac{\langle \psi
_z|h_0-\alpha |\psi _z\rangle }{E_0-\alpha }
\end{equation}
a new K\"ahler potential. It then follows the expression for a new Hermit metric
$$
g^{(1)}=f_{z\overline{z}}^{\left( 1\right) }=\frac{2k+1}{\left( 1-z\overline{%
z}\right) ^2}-\frac{p(2k+p)}{(2k+p-pz\overline{z})^2}.
$$
The unit disk represents now a phase space with the varying Gauss
curvature ${\cal K}^{(1)}=-\frac 2g\frac{\partial ^2}{\partial z\partial
\overline{z}}\ln g^{(1)}.$ The explicit expression for ${\cal K}^{(1)}$ is
rather complicated so that be omitted here. We will note only that ${\cal K%
}^{(1)}\rightarrow 0$ when $k\rightarrow \infty $.

The function
$$
\delta ^{\left( 1\right) }\left( z,\overline{z}^{\prime }\right)
=\sum_{n=0}^\infty \varphi _n\left( z\right) \overline{\varphi }_n\left(
z^{\prime }\right) =\left( 1-z\overline{z}^{\prime }\right) ^{-2k-1}\frac{%
2k+p-pz\overline{z}}{2k+p}
$$
plays the role of the transformed Bergman kernel which takes off the
integration with respect to the new inner product (\ref{nsc}).

By making use of the covariant Berezin symbols we can establish correspondence between
the quantum mechanical operators and the classical observables which are the
functions of a point in the unit disk. The function
$$
H_1=\langle \varphi _{\overline{z}}\left| h_1\right| \varphi _{\overline{z}%
}\rangle =2k+\frac{4kz\overline{z}\left( 2k+p+1-pz\overline{z}\right) }{%
\left( 1-z\overline{z}\right) \left( 2k+p-pz\overline{z}\right) }
$$
corresponds to the transformed Hamiltonian operator and therefore it is the
transformed Hamilton function. Similarly, the functions $P_{+}$ and $P_{\_}$,
$$
P_{+}=\overline{P}_{-}=\langle \varphi _{\overline{z}}\left| p_{+}\right|
\varphi _{\overline{z}}\rangle =
$$
$$
\frac{4k\overline{z}}{2k+p-pz\overline{z}}\left[ \frac{\left( 2k+1\right)
\left( 2k+2\right) }{\left( 1-z\overline{z}\right) ^2}+\left( p-1\right)
\frac{2k+1}{1-z\overline{z}}+p\left( p-1\right) \right] ,
$$
correspond to the lowering and raising operators.

The change of the K\"ahler potential induces the change of the symplectic
form which remains closed because of its dimension. As a consequence we
have another Poisson brackets in the unit disk. The functions $P_0=\frac 12H_1$
and $P_{\pm }$ form now neither a closed Lie algebra nor a polynomial one with
respect to the new Poisson bracket but it is not difficult to check the
following relations:
$$
\left\{ P_0,P_{\pm }\right\} =\pm iP_{\pm },\quad \left\{ P_0,z\right\}
=iz,\quad \left\{ P_0,\overline{z}\right\} =-i\overline{z}.
$$
Two latter relations mean that the equation of trajectories for the transformed
classical system in the transformed phase space remains unchanged under the Darboux transformation.

\vspace{5mm}
\begin{flushleft}
{\bf VI. CONCLUSION}
\end{flushleft}

\vspace{5mm}
Using the coherent states technique we have constructed a classical counterpart
of the transformed $a$ $l\acute a$ Darboux quantum singular oscillator. It has been
shown that the Darboux transformation for this system translates into the
transformation of the K\"ahler potential. New K\"ahler potential has a
supplementary summand of the form $\ln (H_0-\alpha )$ where $H_0$ is initial
Hamilton function corresponding to the quantum Hamiltonian operator for
the singular oscillator. This leads to the distortion of the initial phase
space.
However, the distortion is consistent with the change of the Hamilton function in such a manner that the curves of a constant energy (classical trajectories) in the transformed phase space remain unchanged. A similar observation has recently been discusse
d in Lagrangian formalism and for the canonical method of quantization.%
\cite{NFilho97}

\vspace{0.5cm}
\begin{flushleft}  {\bf ACKNOWLEDGMENT}
\end{flushleft}
This research has been partially supported by RFBR grant No 97-02-16279.

\end{document}